\documentstyle[11pt,newpasp,twoside,epsf,epsfig]{article}
\def\edcomment#1{\iffalse\marginpar{\raggedright\sl#1\/}\else\relax\fi}
\reversemarginpar

\begin{document}

\def\etal{{\it et~al.~}}
\def\bsax{{\it BeppoSAX~}}
\def\ginga{{\it Ginga~}}
\def\astroe{{\it ASTRO-E~}}
\def\constellation{{\it CONSTELLATION~}}
\def\einstein{{\it Einstein~}}
\def\exosat{{\it EXOSAT~}}
\def\tenma{{\it Tenma~}}
\def\asca{{\it ASCA~}}
\def\rosat{{\it ROSAT~}}
\def\rxte{{\it RXTE~}}
\def\euve{{\it EUVE~}}
\def\heao1{{\it HEAO-1~}}
\def\integral{{\it INTEGRAL~}}
\def\cl{clusters of galaxies~}
\def\nt{nonthermal~}
\def\fov{field of view~}
\def\chandra{{\it Chandra~}}
\def\xmm{{\it XMM-Newton~}}
\def\kbeta{$K_{\beta}$~}
\def\kalpha{$K_{\alpha}$~}
\def\pho{~{\rm ph~cm}^{-2}~{\rm s}^{-1}~{\rm keV}^{-1}~}
\def\erg{~{\rm erg~ cm}^{-2}\ {\rm s}^{-1}~}
\def\ergs{~{\rm erg~s}^{-1}~}
\def\h0{~{\rm H_0 = 50~km~s}^{-1}\ {\rm Mpc}^{-1}~h_{50}~}

\title{High Energy Results from BeppoSAX}
\author{Roberto Fusco-Femiano}
\affil{IASF/CNR, Roma, Italy} \author{Daniele Dal Fiume}
\affil{TESRE/CNR, Bologna, Italy} \author{Mauro
Orlandini}\affil{IASF/CNR, Bologna, Italy}\author{Sabrina De
Grandi}\affil{Osservatorio di Merate, Merate, Italy}\author{
Silvano Molendi}\affil{IASF/CNR, Milano, Italy}\author{Luigina
Feretti}\affil{IRA/CNR, Bologna, Italy}\author{Paola
Grandi}\affil{IASF/CNR, Roma, Italy}\author{Gabriele Giovannini}
\affil{Universita` di Bologna, Bologna, Italy}

\begin{abstract}
We review all the \bsax results relative to the search for
additional \nt components in the spectra of clusters of galaxies.
In particular, our MECS data analysis of A2199 does not confirm
the presence of the \nt excess reported by Kaastra \etal (1999). A
new observation of A2256 seems to indicate quite definitely that
the \nt fluxes detected in Coma and A2256 are due to a diffuse \nt
mechanism involving the intracluster medium. We report marginal
evidence ($\sim 3\sigma$) for a \nt excess in A754 and A119, but
the presence of point sources in the field of view of the PDS
makes unlikely a diffuse interpretation.
\end{abstract}

\section{Introduction}

It is well known that X-ray measurements in the energy range 1-10
keV of thermal bremsstrahlung emission from the hot, relatively
dense intracluster gas, have already contributed in an essential
way to our understanding of the cluster environment. However,
recent researches on \cl have unveiled new spectral components in
the intracluster medium (ICM) of some clusters, namely a cluster
soft excess discovered by \euve (Lieu \etal 1966) and a hard X-ray
(HXR) excess detected by \bsax (Fusco-Femiano \etal 1999) and
\rxte (Rephaeli, Gruber, \& Blanco 1999). Observations at low and
high energies can give additional insights on the physical
conditions of the ICM.

Nonthermal emission was predicted at the end of seventies in \cl
showing extended radio emission, radio halos or relics (see
Rephaeli 1979). In particular, the same radio synchrotron
electrons can interact with the CMB photons to give inverse
Compton (IC) nonthermal X-ray radiation. Attempts to detect \nt
emission from a few \cl were performed with balloon experiments
(Bazzano \etal 1984;90), with \heao1 (Rephaeli, Gruber \&
Rothschild 1987; Rephaeli \& Gruber 1988), with the OSSE
experiment onboard the \textit{Compton-GRO} satellite (Rephaeli,
Ulmer \& Gruber 1994) and with \rxte \& \asca (Delzer \& Henriksen
1998), but all these experiments reported essentially flux upper
limits. However, we want to remind the conclusions of the paper
regarding the OSSE observation of HXR radiation in the Coma
cluster by Rephaeli, Ulmer \& Gruber in 1994: "\textit{..It can be
definitely concluded that the detection of the IC HEX (high energy
X-ray) emission necessitates an overall sensitivity a few times
$10^{-6}\pho$ in the 40-80 keV band. ..To reduce source confusion,
detectors optimized specifically for HEX measurements of clusters
should have $\sim 1^{\circ}\times 1^{\circ}$ fields of view. A
level of internal background more than a factor of 10 lower than
that of OSSE is quite realistic. Obviously, another very desirable
feature of any future experiment is wide energy coverage, starting
near (or below) 15-20 keV, in order to independently measure the
tail of the thermal emission}". In these conclusions it is
possible to find the spectral characteristics of the Phoswich
Detector System (PDS) onboard \bsax which is able to detect hard
X-ray emission in the 15-200 keV energy range. The PDS uses the
rocking collimator technique for background subtraction with angle
of 3.5$^{\circ}$. The strategy is to observe the X-ray source with
one collimator and to monitor the background level on both sides
of the source position with the other in order to have a
continuous monitoring of the source and background. The dwell time
is 96 sec. The background level of the PDS is the lowest obtained
so far with high-energy instruments onboard satellities ($\sim
2\times 10^{-4} ~\rm {counts~s}^{-1}~\rm keV^{-1}~$ in the 15-200
KeV energy band) thanks to the equatorial orbit of \bsax. The
background is very stable again thanks to the favorable orbit, and
no modelling of the time variation of the background is required
(Frontera \etal 1997).

\section{Hard X-ray observations of clusters of galaxies by BeppoSAX}

\bsax observed seven clusters of galaxies with the main objective
to detect additional nonthermal components in their spectra.

\subsection{Coma}
The first cluster was Coma observed in December 1997 for an
exposure time of about 91 ksec. A nonthermal excess with respect
to the thermal emission was observed (Fusco-Femiano \etal 1999) at
a confidence level of about 4.5$\sigma$ (see Fig.1).
\begin{figure}
\centerline{\psfig{figure=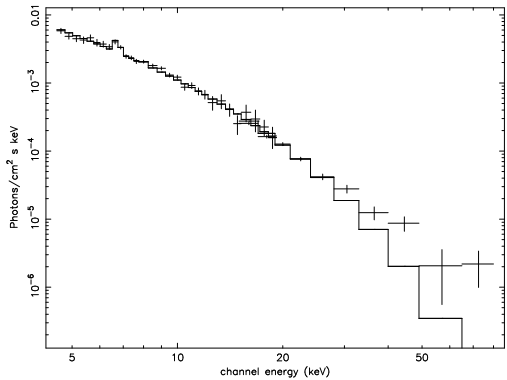,height=3.0in,angle=-90}}
\caption{Coma cluster - HPGSPC and PDS data. The continuous line
represents the best fit with a thermal component at the average
cluster gas temperature of 8.5$^{+0.6}_{-0.5}$ keV.
\label{coma}}
\end{figure}
The thermal emission was measured with the HPGSPC always onboard
\bsax in the 4-20 keV energy range with a FWHM ($\sim
1^{\circ}\times 1^{\circ}$) comparable to that of the PDS ($\sim
1.3^{\circ}$, hexagonal). The average gas temperature is
$8.5^{+0.6}_{-0.5}$ keV consistent with the temperature of Ginga
of 8.2 keV (David \etal 1993). The $\chi^2$ value has a
significant decrement when a second component, a power law, is
added. On the other hand, if we consider a second thermal
component, instead of the \nt one, the fit requires a temperature
greater than 40-50 keV. This unrealistic value may be interpreted
as a strong indication that the detected hard excess is due to a
\nt mechanism. The data are not able to give a good determination
of the photon spectral index (0.7-2.5; 90\%), but the \nt flux
$\sim 2.2\times 10^{-11}\erg$ in the 20-80 keV energy range is
rather stable against variations of the power-law index. Binning
the PDS data between 40-80 keV the \nt flux is lower by a factor
about 2 with respect to the upper limit derived by the OSSE
experiment (see Fig. 1 of Fusco-Femiano \etal 1999). In the same
time a \rxte observation of the Coma cluster (Rephaeli, Gruber \&
Blanco 1999) showed evidence for the presence of a second
component in the spectrum of this cluster, in particular the
authors argued that this component is more likely to be
nonthermal, rather than a second thermal component at lower
temperature.

The first possible explanation for the detected excess is emission
by a point source in the \fov of the PDS. The most qualified
candidate is X Comae, a Seyfert 1 galaxy (z=0.092). \rosat PSPC,
\exosat and \einstein IPC observations report a flux at
approximately the same flux level of $1.6\times 10^{-12}\erg$ in
the 2-10 keV energy band. With a typical photon index of 1.8 the
variability factor of the source to account for the detected
excess is of the order of 10 which could be still plausible. But
luckily enough, X Comae is located just on the edge of the \fov of
the MECS (see Fig. 3 of Fusco-Femiano 1999). Considering the
location of X Comae and the lack of detection, it is possible to
estimate an upper limit to the flux of the source of $\sim 4\times
10^{-12}\erg$ (2-10 keV) when \bsax observed Coma which is a
factor $\sim$ 7 lower than the flux of $\sim 2.9\times
10^{-11}\erg$ required to account for the \nt HXR emission in the
PDS. A recent mosaic of the Coma cluster with \xmm (Briel \etal
2000) reports a tentative identification of 3 quasars in the
central region, but the estimated fluxes are insufficient to
reproduce the excess detected by \bsax. However, we cannot exclude
that an obscured source, like Circinus (Matt \etal 1999) a Seyfert
2 galaxy very active at high X-ray energies, may be present in the
\fov of the PDS. With the MECS image it is possible to exclude the
presence of this kind of source only in the central region of
about 30$'$ in radius unless the obscured source is within 2$'$ of
the central bright core. We have estimated that the probability to
find an obscured source in the \fov of the PDS is of the order of
10\% and also Kaastra \etal (1999) independently arrived to the
same estimate.

Another interpretation is that the \nt emission is due to
relativistic electrons scattering the CMB photons and in
particular the same electrons responsible for the radio halo
emission present in the central region of the cluster, Coma C. In
this case we can derive the volume-averaged intracluster magnetic
field, $B_X$, using only observables, combining the X-ray and
radio data (see Eq. 1 of Fusco-Femiano \etal 1999). The value of
$B_X$ is of the order of 0.15 $\mu$G and assuming a radio halo
size of R = 1 Mpc at the distance of Coma also the electron energy
density ($\sim 7\times 10^{-14}$ erg cm$^{-3}$) can be derived.
The value of the magnetic field derived by the \bsax observation
seems to be inconsistent with the measurements of Faraday rotation
of polarized radiation of sources through the hot ICM that give a
line-of-sight value of $B_{FR}$ of the order of 2-6 $\mu$G (Kim
\etal 1990; Feretti \etal 1995). But Feretti and collaborators
inferred also the existence of a weaker magnetic field component,
ordered on a scale of about a cluster core radius, with a
line-of-sight strength in the range 0.1-0.2 $\mu$G consistent with
the value derived from \bsax. So, we can argue that the component
at 6 $\mu$G is likely present in local cluster regions, while the
overall cluster magnetic field may be reasonably represented by
the weaker and ordered component. However, there are still many
and large uncertainties on the value of the magnetic field
determined using the FR measurements (Newman, Newman, \& Rephaeli
2002). Other determinations of B based on different methods are in
the range 0.2-0.4 $\mu$G (Hwang 1997; Bowyer \& Bergh\"{o}fer
1998; Sreekumar \etal 1996; Henriksen 1998). The equipartition
value is of the order of $\sim 0.4\mu$G (Giovannini \etal 1993).

However, alternative interpretations to the IC model have been
proposed essentially motivated by the discrepancy between the
values of $B_X$ and $B_{FR}$. Blasi \& Colafrancesco (1999) have
suggested a secondary electron production due to cosmic rays
interactions in the ICM. However, this model implies a
$\gamma$-ray flux larger than the EGRET upper limit, unless the
hard excess and the radio halo emission are due to different
populations of electrons. A different mechanism is given by \nt
bremsstrahlung from suprathermal electrons formed through the
current acceleration of the thermal gas (Ensslin, Lieu, \&
Biermann 1999; Dogiel 2000; Sarazin \& Kempner 2000; Blasi 2000).
At present, due to the low efficiency of the proposed acceleration
processes and of the bremsstrahlung mechanism, these models would
require an unrealistically high energy input, as pointed out by
Petrosian (2001). Regarding the discrepancy between $B_X$ and
$B_{FR}$, Goldshmidt \& Rephaeli already in 1993 suggested that
this discrepancy could be alleviated by taking into consideration
the expected spatial profiles of the magnetic field and
relativistic electrons. More recently, it has been shown that IC
models that include the effects of more realistic electron
spectra, combined with the expected spatial profiles of the
magnetic fields, and anisotropies in the pitch angle distribution
of the electrons allow higher values of the intracluster magnetic
field , in better agreement with the FR measurements (Brunetti
\etal 2001; Petrosian 2001).

\subsection{A2199}

The cluster has been observed in April 1997 for 100 ksec (Kaastra
\etal 1999). The MECS data in the range 8-10 keV seem to show the
presence of a hard excess with respect to the thermal emission.
Between 9$'$ and 24$'$, the count rate is 5.4$\pm$0.6 counts
ks$^{-1}$, while the best fit thermal model predicts only 3.4
counts ks$^{-1}$. So, the excess is at a confidence level of $\sim
3.3\sigma$. The PDS data are instead not sufficient to prove the
existence of a hard tail. There are some difficulties to account
for the presence of a \nt excess in this cluster because the
electrons responsible for the hard emission would have an energy
of $\sim$ 4 GeV and a resulting IC lifetime of only $\sim 3\times
10^{8}$ yr. So, these electrons have to be replenished by a
continuous acceleration process and this is particularly difficult
to explain in A2199 that is a bright cooling flow cluster, a
regular cluster without merger events able to release a fraction
of the input energy in particle acceleration. However, a source of
relativistic electrons may be given via the decay of pions
produced by proton-proton collisions between intracluster cosmic
rays and gas, as suggested by Blasi \& Colafrancesco (1999). We
have re-analyzed the MECS data and Fig. 2 shows only a point above
the thermal model at the level of $\sim 2\sigma$.
\begin{figure}
\centerline{\psfig{figure=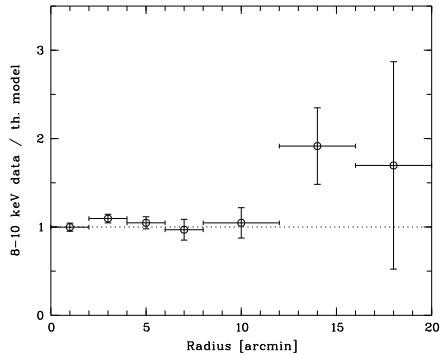,height=3.0in,angle=-90}}
\caption{A2199 - MECS data. The points represent the ratio of the
data above the MEKAL model in the energy range 8-10 keV.
\label{fig:A2199}}
\end{figure}
However, the cluster is planned to be observed by \xmm that should
be able to discriminate between these two different results of the
MECS data analysis, considering the low average gas temperature of
about 4.5 keV (David \etal 1993).

\subsection{A2256}

The cluster A2256 is the second cluster where \bsax detected a
clear excess (see Fig. 3) at about 4.6$\sigma$ above the thermal
emission (Fusco-Femiano \etal 2000).
\begin{figure}
\centerline{\psfig{figure=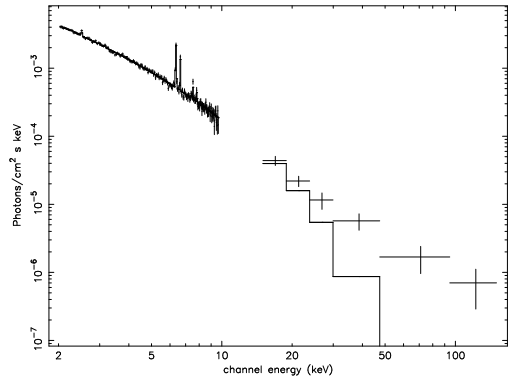,height=3.0in,angle=-90}}
\caption{Abell 2256 - MECS and PDS data. The continuous line
represents the best fit with a thermal component at the average
cluster gas temperature of 7.47$\pm$0.35 keV. \label{fig:A2256}}
\end{figure}
The temperature is about 7.4$\pm$0.23 keV consistent with the
value determined by previous observations of \asca, \einstein and
\ginga. The thermal emission is measured by the MECS taking into
account the difference between the two fields of view of the two
instruments. Also in this case the $\chi^2$ value has a
significant decrement when a second component, a power law, is
added and also in this case the fit with a second thermal
component gives an unrealistic temperature which can be
interpreted as evidence in favour of a \nt mechanism for the
second component present in the X-ray spectrum of A2256. The range
of the photon index at 90$\%$ confidence level is very large :
0.3-1.7. The flux of the \nt component is $\sim 1.2\times
10^{-11}\erg$ in the 20-80 keV energy range, rather stable against
variations of the photon index.

There is only a QSO in the \fov of the PDS, QSO 4C+79.16, observed
by \rosat with a count rate of $\sim$ 0.041 c/s and about 1.2 c/s
are necessary to reproduce the observed \nt excess and considering
that the QSO is $\sim 52'$ off-axis an unusual variability of
about 2 orders of magnitude is required. The MECS image excludes
the presence of an obscured source in the central region ($\sim
30'$ in radius) of the cluster. We want to stress that the
analysis of A2256 regards two observations (46 \& 96 ksec) with a
time interval of $\sim$ 1 yr (Feb. 98 and Feb. 99) and this
analysis does not show significant flux variations. In addition,
we have re-observed the cluster after about two years from the
previous one and the two spectra are consistent (see Fig. 4) and
also in this case the observation is composed of two observations
with a time interval of $\sim$ 1 month and the analysis does not
show significant flux variations.
\begin{figure}
\centerline{\psfig{figure=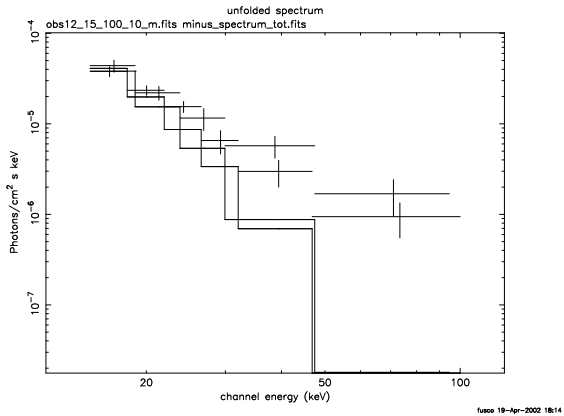,height=3.0in,angle=-90}}
\caption{A2256 - PDS data of two observations (Feb. 98/Feb. 99 and
July 01/August 01). The continuous lines represent the best fit to
the two data sets with a thermal component at the average cluster
gas temperature of 7.47$\pm$0.35 keV; the error bars are at
1$\sigma$. \label{fig:a2256NO}}
\end{figure}
So, this results and the fact that the two clusters with a
detected hard excess, Coma and A2256, both have extended radio
emission, make less plausible the point source interpretation and
strongly support the idea of a diffuse nonthermal mechanism
involving the ICM.

The diffuse radio emission of A2256 is very complex. It is
composed of a relic at a distance of about 8$'$ from the center. A
broad region (1$\times$0.3 Mpc) with a rather uniform and flat
spectral index of 0.8$\pm$0.1 between 610 and 1415 MHz (Bridle
\etal 1979). There is a second fainter extended component in the
cluster center with a steeper radio spectral index of $\sim$ 1.8
(Bridle \etal 1979, Rengelink \etal 1997). Markevitch \& Vikhlinin
(1997) in their analysis of the \asca data noted a second
component in the spectrum of A2256 in the central spherical bin of
radius 3$'$. Their best fit is a power law with a photon index of
2.4$\pm$0.3 which therefore favors a \nt component. Considering
that there are no bright point sources in the \rosat HRI image
they suggested the presence of an extended source. Also the joint
analysis \asca GIS \& \rxte PCA (Henriksen 1999) is consistent
with a detection of a nonthermal component in addition to a
thermal component. The MECS data do not show this steep \nt
component in the central bin because the energy range is truncated
to a lower limit of 2 keV (Molendi, De Grandi \& Fusco-Femiano
2000). So, in conclusion the power law with slope 2.4 found in the
\asca data and the upper limit 1.7 determined by \bsax suggest
that two tails could be present in the X-ray spectrum of A2256.
The former might be due to the radio halo with the steep index of
1.8 that is not visible in the PDS (we estimate a flux of $\sim
4\times 10^{-13}\erg$), and the latter might be due to the relic
with a flatter radio index of 0.8$\pm$0.1 that indicates a broad
region of reaccelerated electrons, probably the result of the
ongoing merger event shown by a \chandra observation (Sun \etal
2002).

\subsection{A1367}

A \bsax observation of Abell 1367 has not detected hard X-ray
emission in the PDS energy range above 15 keV (P.I.: Y.Rephaeli).
A1367 is a near cluster (z=0.0215) that shows a relic at a
distance of about 22$'$ from the center and a low gas temperature
of $\sim$ 3.7 keV (David \etal 1993) that might explain lack of
thermal emission at energies above 15 keV. We do not expect
presence of \nt radiation for two reasons : the radio spectral
index $\alpha_R$ = 1.90$\pm$0.27 (Gavazzi \& Trinchieri 1983)
seems to indicate the absence of high energy reaccelerated
electrons and in any case the steep spectrum gives a negligible
flux in the PDS. Besides, the radio region has a limited extent of
8$'$ corresponding to 300 kpc. The source has been observed also
by \xmm and the data analysis is still in progress.

\subsection{A3667}

A3667 is one of the most spectacular \cl. It contains one of the
largest radio sources in the southern sky with a total extent of
about 30$'$ which corresponds to about 2.6$h_{50}^{-1}$ Mpc. A
similar but weaker region is present also to the south-west
(Robertson 1991; R\"{o}ttgering \etal 1997). The Mpc-scale radio
relics may be originated by the ongoing merger visible in the
optical region, in the X-ray, as shown by the elongated isophotes,
and in the weak lensing map. The \asca observation reports an
average gas temperature of 7.0$\pm$0.6 (Markevitch, Sarazin, \&
Vikhlinin 1999). The temperature map shows that the hottest region
is in between the two groups of galaxies confirming the merger
scenario. The PDS \fov includes only the radio region in the north
of the cluster. A long observation with the PDS (effective
exposure time 44+69 ksec) reports a clear detection of hard X-ray
emission up to about 35 keV at a confidence level of $\sim
10\sigma$. Instead, the fit with a thermal component at the
average gas temperature indicates an upper limit for the \nt flux
of $\sim 6.4\times 10^{-12}\erg$ in the 20-80 keV energy range
that is a factor $\sim$ 3.4 and $\sim$ 2 lower than the \nt fluxes
detected in Coma and A2256, respectively (see Fig. 5). In the IC
interpretation this flux upper limit combined with the radio
synchrotron emission determines a lower limit to the
volume-averaged intracluster magnetic field of 0.41$\mu$G.

\begin{figure}
\centerline{\psfig{figure=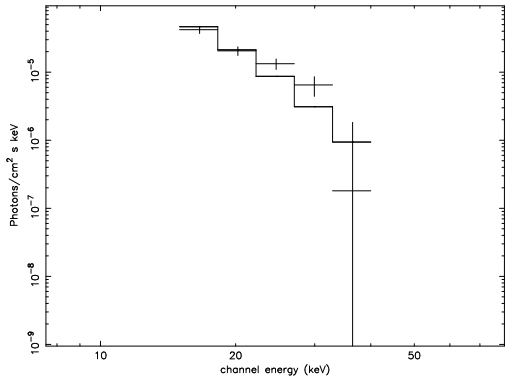,height=3.0in,angle=-90}}
\caption{A3667 - PDS data. The continuous lines represent the best
fit with a thermal component at the average cluster gas
temperature of 7.0$\pm$0.6 keV. \label{fig:a3667}}
\end{figure}

Given the presence of such a large radio region in the NW of the
cluster, a robust detection of a \nt component might be expected
instead of the upper limit reported by \bsax. One possible
explanation may be related to the radio spectral structure of the
NW relic. The sharp edge of the radio source ($\alpha_R\sim 0.5$)
is the site of particle acceleration (Roettiger, Burns \& Stone
1999) , while the progressive index steepening with the increase
of the distance from the shock ($\alpha_R\sim 1.5$) would indicate
particle ageing because of radiative losses. In the narrow shocked
region, where particle reacceleration is at work, the magnetic
field is expected to be amplified by adiabatic compression with
the consequence that the synchrotron emission is enhanced thus
giving a limited number of electrons able to produce IC X-rays. In
the post-shock region of the relic the electrons suffer strong
radiative losses with no reacceleration, considering also that the
relic is well outside the cluster core. Therefore, the electron
energy spectrum develops a high energy cutoff at $\gamma < 10^4$
and the electron energy is not sufficiently high to emit IC
radiation in the hard X-ray band. Synchrotron emission is detected
from the post-shocked region, because the magnetic field is still
strong enough due to the likely long time to relax.

\subsection{A754 \& A119}

The last two clusters, A754 and A119, observed by \bsax show an
evident merger activity. It is plausible that a considerable
fraction of the input energy during a merger process can be
released in particle acceleration and remitted in various energy
bands. The scope of these observations was to verify whether
clusters showing merger events can produce \nt X-ray radiation
also in the absence of a clear evidence of diffuse radio emission
as it is for Coma and A2256.

The rich and hot cluster A754 is considered the prototype of a
merging cluster. X-ray observations report a violent merger event
in this cluster (Henry \& Briel 1995; Henriksen \& Markevitch
1996; De Grandi \& Molendi 2001), probably a very recent merger as
shown by a numerical hydro/N-body model (Roettiger, Stone, \&
Mushotzky 1998). Therefore, the intracluster medium of A754
appears to be a suitable place for the formation of radio halos or
relics. As a consequence, radio and HXR observations of this
cluster are relevant to verify the suggested link between the
presence of nonthermal phenomena and merger activity in clusters
of galaxies. The cluster has been recently observed with the NRAO
VLA observatory (Kassim \etal 2001), after our \bsax proposal,
suggesting the existence of a radio halo and at least one radio
relic. The presence of a radio halo is confirmed by a deeper
observation at higher resolution (Fusco-Femiano \etal, in
preparation). A754 was observed in hard X-rays with \rxte in order
to search for a nonthermal component (Valinia \etal 1999) and the
fit to the PCA and HEXTE data set an upper limit of $\sim$
1.4$\times$ 10$^{-12}\erg$ in the 10-40 keV band to the nonthermal
emission.

A long \bsax observation of A754 shows an excess at energies above
about 45 keV with respect to the thermal emission at the
temperature of 9.4 keV (see Fig. 6).
\begin{figure}
\centerline{\psfig{figure=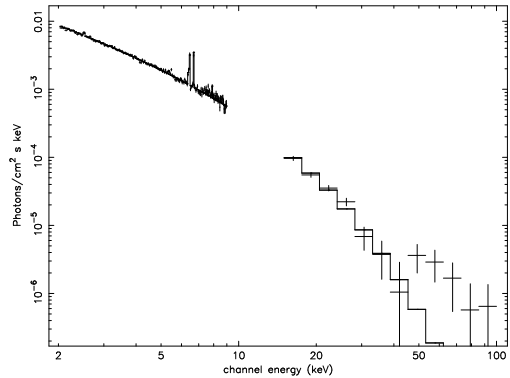,height=3.0in,angle=-90}}
\caption{A754 - MECS and PDS data. The continuous lines represent
the best fit with a thermal component at the average cluster gas
temperature of $9.42^{+0.16}_{-0.17}$ keV\label{fig:a754}}
\end{figure}
The excess is at a level of confidence of 3$\sigma$. The \nt flux
is $\sim 1\times 10^{-11}\erg$ in the range 40-100 keV consistent
with the flux upper limit determined by \rxte ($\sim 1.6\times
10^{-12}\erg$ in the range 10-40 keV).

There are two possible origins for the detected excess. One is
tied to the presence of the diffuse radio emission and the other
explanation is due to the presence of the radio galaxy 26W20 in
the \fov of the PDS discovered by Harris \etal (Westerbork radio
survey, Harris \etal 1980). This source shows X-ray
characteristics similar to those of a BL Lac object. The radio
galaxy has had several X-ray observations, due to its close
proximity to A754, and all these observations give a flux of $\sim
2.3\times 10^{-12}\erg$ in the 0.5-3 keV energy range. The source
shows variability (18\% in 5 days in 1992).
\begin{figure}
\centerline{\psfig{figure=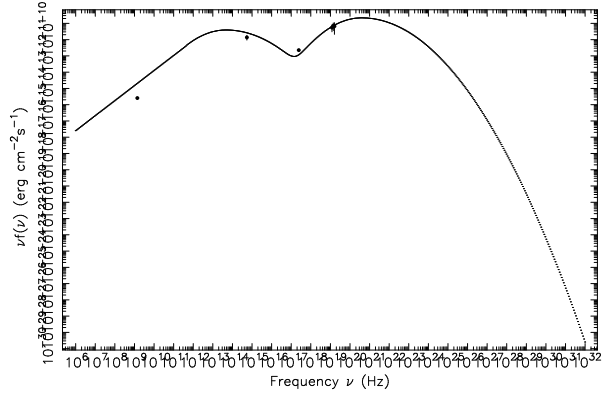,height=4.0in,angle=-90}}
\caption{Spectral energy distribution for 26W20. The highest
energy points  refer to the PDS observation. The dotted line is
the fit to the SED.\label{fig:sed}}
\end{figure}
The fit to the SED for 26W20 (see Fig. 7), where the highest
energy points refer to the PDS observation assuming that this
source is responsible for the detected excess, requires a flat
index of about 0.3 to extrapolate the flux detected by \rosat in
the PDS energy range, taking into account the angular response of
the detector. Unfortunately, the source is not in the field of
view of the MECS because it is hidden by one of the calibration
sources of the instrument. The conclusion is that a HXR
observation with spatial resolution is necessary to discriminate
between these two interpretations.

Finally, A119 was the last cluster observed by \bsax to detect an
additional \nt component in its X-ray spectrum. \rosat PSPC, \asca
and \bsax observations have shown a rather irregular and
asymmetric X-ray brightness suggesting that the cluster is not
completely relaxed and may have undergone a recent merger
(Cirimele \etal 1997; Markevitch \etal 1998; Irwin, Bregman \&
Evrard 1999; De Grandi \& Molendi 2001). The average cluster
temperature measured by \bsax is 5.66$\pm$0.16 keV within 20$'$
and is consistent with previous measurements of \einstein and
\exosat. The excess with respect to the thermal emission at the
average gas temperature measured by the MECS is at confidence
level of $\sim 2.8\sigma$ (see Fig. 8).
\begin{figure}
\centerline{\psfig{figure=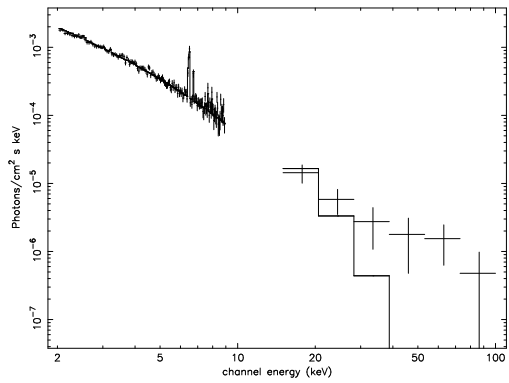,height=3.0in,angle=-90}}
\caption{A119 - MECS and PDS data. The continuous lines represent
the best fit with a thermal temperature at the avergage cluster
gas temperature of $5.66\pm 0.16$ keV.\label{fig:a119}}
\end{figure}
The \nt flux is in the range $7-8\times 10^{-12}\erg$ in the 20-80
keV energy range and $3-4\times 10^{-12}\erg$ in the 2-10 keV
energy band for a photon spectral index in the range 1.5-1.8.

A119 does not show evidence of a radio halo or relic, but the
presence of a recent merger event could accelerate particles able
to emit nonthermal emission in the PDS energy range. However, the
presence of 7 QSO with redshift in the range 0.14-0.58 makes very
unlikely that this possible excess at a flux level of $3-4\times
10^{-12}\erg$ in the 2-10 keV energy band may be due to a diffuse
source. We can instead exclude that the observed excess is due to
the radio source 3C29, a FR I source located in the \fov of the
MECS at a distance of about 21$'$ from the \bsax pointing.

\begin{figure}
\centerline{\psfig{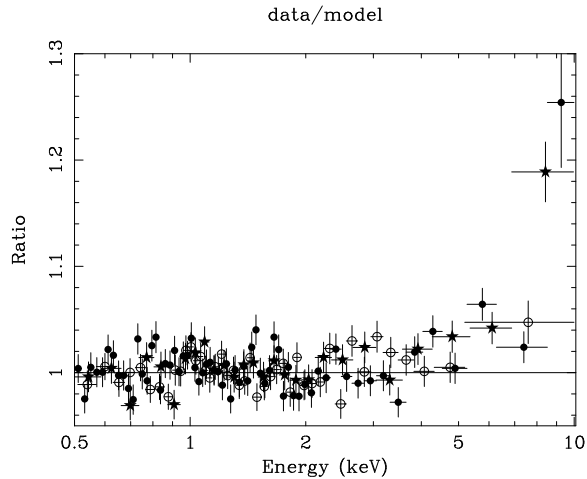}}
\caption{A2256 - Residuals in the form of a ratio of data to a
thermal MEKAL model. The best fit temperature for the simulated
spectrum is $\sim$ 4 keV. Full circles and stars are for the PN
single and double events spectra, and open circles are for the MOS
spectrum. \label{fig:a2256_sim}}
\end{figure}

\section{Conclusions}

\bsax observed a clear evidence of \nt emission in two clusters,
Coma and A2256, both showing extended radio regions. In
particular, the two observations of A2256 strongly support the
presence of a diffuse non thermal mechanism involving the ICM.
These detections and the lack of detection in other clusters seem
to indicate that the essential requirement to observe additional
\nt components at the level of the PDS sensitivity is the presence
of large regions of reaccelerated electrons, with Lorenz factor
$10^4$, due to the balance between radiative losses and
reacceleration gains in turbulence generated by merger events that
must be very recent considering the short lifetime of the
electrons.

\bsax, as it is well known, has ceased its activity at the end of
April 2002. The next missions able to search for \nt components
are :
\par\noindent
$\bullet$ \integral. In particular, with IBIS, that has a spatial
resolution of 12$'$, we have the opportunity a) to localize the
source of the nonthermal X-ray emission. In the case of a point
source it is possible to identify it, while in the case of a
diffuse source it is possible to verify whether the \nt emission
is mainly concentrated in the cluster central region or in the
external region, as predicted by the model for the Coma cluster of
Brunetti \etal (2001), or it is uniformly spread over the whole
radio halo present in the cluster. b) to have a better
determination of the photon spectral index.
\par\noindent
$\bullet$ \astroe. The Hard X-ray Detector (HXD) has a \fov of
$34'\times 34'$ similar to that of the MECS. A positive detection
of the \nt emission already measured by \bsax in Coma and A2256
would eliminate the ambiguity between a diffuse emission involving
the intracluster gas and a point source, considering that the MECS
images do not show evidence for point sources.
\par\noindent
$\bullet$ The future missions are represented by {\it NEXT} and
\constellation.

\vskip5pt

These missions will be operative in the next years, but the energy
range and the spectral capabilities of \xmm/EPIC give the
possibility to localize \nt components in regions of low gas
temperature as shown by the simulation regarding the radio relic
of A2256 performed using the \nt flux measured by \bsax (see Fig.
9). This region has a gas temperature of 4 keV likely associated
with the ongoing merger shown by a \chandra observation (Sun \etal
2001). So with \xmm we should have the possibility, by comparing
the X-ray and radio structures, to constrain the profiles of the
magnetic field and of relativistic electrons.


\end{document}